# Giant Magnetochiral Anisotropy in Weyl-semimetal $WTe_2$ Induced by Diverging Berry Curvature

Tomoyuki Yokouchi[1†], Yuya Ikeda[2], Takahiro Morimoto[2], and Yuki Shiomi[1]

[1]*Department of Basic Science, The University of Tokyo, Tokyo 152-8902, Japan*

[2]*Department of Applied Physics, The University of Tokyo, Tokyo 113-8656, Japan*

[†] To whom correspondence should be addressed. E-mail: yokouchi@g.ecc.u-tokyo.ac.jp

**Abstract**

The concept of Berry curvature is essential for various transport phenomena. However, an effect of the Berry curvature on magnetochiral anisotropy, i.e. nonreciprocal magneto-transport, is still elusive. Here, we report the Berry curvature originates the large magnetochiral anisotropy. In Weyl-semimetal $WTe_2$, we observed the strong enhancement of the magnetochiral anisotropy when the Fermi level is located near the Weyl points. Notably, the maximal figure of merit $\bar{\gamma}$ reaches $1.2\times10^{-6}$ $m^2T^{-1}A^{-1}$, which is the largest ever reported in bulk materials. Our semiclassical calculation shows that the diverging Berry curvature at the Weyl points strongly enhances the magnetochiral anisotropy.

Whether physical responses are allowed or prohibited is closely related to symmetries [1]. For example, in the linear response regime, the anomalous Hall effect is allowed when time reversal symmetry is broken [2]. In contrast, spin Hall effect is allowed independent of both time reversal and crystalline symmetries [2]. In the nonlinear response regime, a nonlinear Hall effect without external magnetic field is allowed in a bilayer $WTe_2$ but prohibited in a bulk $WTe_2$ due to the difference in the crystalline symmetries between them [3,4]. In addition, the bulk photovoltaic effect, a nonlinear optical phenomenon, is allowed in system without spatial inversion symmetry [4, 5, 6]. Beyond the symmetry argument, the microscopic mechanisms of these physical phenomena have been intensively studied. An important finding is that the notion of the Berry curvature of electronic wave function is essential for the description of these phenomena [8]. For example, the intrinsic anomalous and spin Hall effects are proportional to the integral of Berry curvature over the Fermi sea [2]. In addition, recently, it is found that the Berry curvature also manifests itself in nonlinear responses; the nonlinear Hall effect under time reversal symmetry is induced by the dipole moment of the Berry curvature [3,4,9,10]. In one of the microscopic mechanisms of the bulk photovoltaic effect (shift current mechanism), photocurrent is quantified by the difference in the Berry connection between valance and conduction bands [11, 12]. Hence, investigating relationship between the Berry curvature and a physical response has been one of the central issues in modern condensed matter physics.

Only when the time reversal and spatial inversion symmetries are simultaneously broken, another class of the nonlinear responses is allowed, that is the magnetochiral anisotropy effect [13, 14, 15, 16, 17, 18, 19, 20, 21, 22, 23, 24, 25, 26, 27, 28, 29, 30, 31] which we focus on here. The magnetochiral anisotropy effect (also called the nonreciprocal magneto-transport effect) is nonreciprocal transport responses triggered by an external magnetic field; under magnetic fields, the longitudinal and transverse resistivity of a material is different for electric current $I$ flowing to the right ($+I$) and to the left ($-I$). The magnetochiral anisotropy effect essentially differs from the nonlinear Hall effect, since the former requires time-reversal symmetry breaking but the latter does not. Hence physical mechanisms different from the nonlinear Hall effect must exist in the magnetochiral anisotropy effect. So far several underlying mechanisms of the magnetochiral anisotropy effect have been clarified. For example, an asymmetric electron scattering [18, 19, 22, 23, 31] and the magnetic-field-induced deformation of the Fermi surface caused by the Zeeman term [17, 20] have been identified as the microscopic mechanisms. From the viewpoint of symmetry, modification of electron motion due to the Berry curvature is also expected to affect the magnetochiral anisotropy effect, since the momentum-integrated Berry curvature can be nonzero in systems with broken time-reversal symmetry. However, the role of the Berry curvature on the magnetochiral anisotropy effect has not been well explored.

To study the effect of Berry curvature on the magnetochiral anisotropy effect, Weyl semimetals with broken spatial inversion symmetry [32, 33] are a suitable candidate. Weyl semimetals are quantum materials characterized by topologically nontrivial band structure and possess pairs of band crossing points termed Weyl points. The Weyl points act as monopoles of the Berry curvature, and the Berry curvature becomes singular and diverges at the Weyl points. Hence, when the Fermi level is located near the Weyl points, the effect of the Berry curvature is maximized and thus the magnetochiral anisotropy effect is expected to be dramatically enhanced. In this paper, we demonstrate this scenario using bulk crystals of a polar Weyl semimetal $WTe_2$. The symmetry of bulk $WTe_2$ is a *Pmn*$2_1$ (polar), and $WTe_2$ possesses four pairs of Weyl points [34,35,36]. Around the Weyl points, the Berry curvature is strongly enhanced as schematically shown in Fig. 1(a).

Generally, the direction of the electric field generated in the magnetochiral anisotropy effect depends on the magnetic field direction [13]. The presence of the mirror plane in $WTe_2$ constrains the expected direction of the voltage drop induced by the magnetochiral anisotropy responses in $WTe_2$ as shown in Figs. 1(b) and (c); the longitudinal response (hereafter called nonreciprocal magnetoresistance) is proportional to $\sin\theta$ and largest when the magnetic field ($H$) is perpendicular to the current. Here, $\theta$ is the relative angle between the magnetic-field direction and the current direction. In contrast, the transverse response (hereafter called second-harmonic Hall effect) is proportional to $\cos\theta$ and largest when $H$ is parallel to the current. Here, we note although the nonreciprocal magnetoresistance was reported in nanometre-thick $WTe_2$ whose Fermi level is apart from the Weyl points [20], both the nonreciprocal magnetoresistance and the second-harmonic Hall effect have not been well investigated when the Femi level is tuned near the Weyl points hosting diverging Berry curvature.

In Figs. 1(d) and (e), we present the magnetic-field dependence of the second-harmonic Hall effect and the nonreciprocal magnetoresistance in bulk $WTe_2$ whose Fermi level is close to the Weyl points (sample F). Here, we evaluate the Fermi level position from the carrier densities of electrons and holes obtained by the Hall resistivity measurement (see [37] for details). Because the magnetochiral anisotropy effect is proportional to the magnitude of the square of the current density $j^2$, we measured the second harmonic longitudinal and transverse resistivity ($\rho_{xx}^{2f}$ and $\rho_{yx}^{2f}$) (see also [37] for details). In accord with the expected contribution from the magnetochiral anisotropy effect, the field profiles of $\rho_{xx}^{2f}$ and $\rho_{yx}^{2f}$ are anti-symmetric against the magnetic field. Here, $\rho_{xx}^{2f}$ and $\rho_{yx}^{2f}$ are anti-symmetrized with respect to the magnetic field (see also [37]). The field profiles of the nonreciprocal magnetoresistance and the second-harmonic Hall effect before anti-symmetrization [$\rho_{xx,\text{meas.}}^{2f}(H)$ and $\rho_{yx,\text{meas.}}^{2f}(H)$] is also anti-symmetric against the magnetic field as shown in Fig. S5 [37]. The figures of merit of the second-harmonic Hall effect $\bar{\gamma}_{yx} = 2\rho_{yx}^{2f}/(\rho_{xx}jB)$ and nonreciprocal magnetoresistance $\bar{\gamma}_{xx} = 2\rho_{xx}^{2f}/(\rho_{xx}jB)$ are as large as $1.2 \times 10^{-6}$ m$^2$T$^{-1}$A$^{-1}$ and $0.34 \times 10^{-6}$ m$^2$T$^{-1}$A$^{-1}$ at 1 T, respectively. Remarkably, $\bar{\gamma}_{yx}$ is approximately 3 times larger than the previously reported largest value of the magnetochiral anisotropy effect in $ZrTe_5$ [29]. Interestingly, the second-harmonic Hall effect (i.e. $\rho_{yx}^{2f}$ for $H // I$) is larger than the nonreciprocal magnetoresistance (i.e. $\rho_{xx}^{2f}$ for $H \perp I$) [Figs. 1(d) and (e)]. This feature cannot be explained by the previously reported asymmetric electron scattering mechanism, in which the magnitude of the second-harmonic Hall effect and the nonreciprocal magnetoresistance must be the same [19]. As will be discussed later, this difference originates from the effects of the Berry curvature and the chiral anomaly.

To further confirm that the observed signals originate from the magnetochiral anisotropy effect, first,

we investigated the dependence of $\rho_{yx}^{2f}$ with $H \mathbin{/\mkern-6mu/} I$ on the input-current frequency ($f$) and on the current density ($j$) in sample A. The Fermi level of sample A is also located near the Weyl points as in sample F. As expected in the second-harmonic Hall effect, the magnitude of $\rho_{yx}^{2f}$ is independent of $f$ and linearly increases with increasing $j$ in the low current density region [Figs. 2(a) and (b)]. Note that the deviation from the linear relationship in Fig. 2(b) at high current is attributed to the temperature increase due to the Joule heating effect; the maximum temperature increase is estimated to be approximately 2 K from the change in the linear longitudinal resistivity. Then, we investigated polarity-direction dependence; the sign of the second-harmonic Hall effect and the nonreciprocal magnetoresistance should be reversed when the polarity direction is reversed [13]. We divided a WTe$_2$ sample into two (samples D and L) so that the polarity directions are opposite to each other [Fig. 2(c)]. As shown in Figs. 2(d)-(g), the signs of $\rho_{yx}^{2f}/\rho_{xx}$ and $\rho_{xx}^{2f}/\rho_{xx}$ are opposite for samples D and L in accord with the expected trends in the second-harmonic Hall effect and the nonreciprocal magnetoresistance. We also measured $\rho_{yx}^{2f}/\rho_{xx}$ as a function of the relative angle ($\theta$) between the magnetic-field direction and the current direction. As shown in Fig. 3(a), $\rho_{yx}^{2f}/\rho_{xx}$ obeys $\cos\theta$ as expected. These results corroborate that the observed signals originate from the magnetochiral anisotropy effect.

Then, we investigate the temperature dependence of the second-harmonic Hall effect and the nonreciprocal magnetoresistance. In Figs. 3(b) and (c), we show the temperature dependence of $\rho_{yx}^{2f}/\rho_{xx}$ and $\rho_{xx}^{2f}/\rho_{xx}$ at 9 T for sample A and sample F. As can be seen from Figs. 3(b) and (c), the absolute values of $\rho_{yx}^{2f}/\rho_{xx}$ and $\rho_{xx}^{2f}/\rho_{xx}$ increase notably with decreasing temperature. The enhancement at low temperatures should be closely related to the microscopic mechanism and will be discussed later.

Next, we investigated the Fermi level dependence of the second-harmonic Hall effect and the nonrecirpcal mangetoresistance. In WTe$_2$, it is known that the Fermi level position changes due to a deficiency of Te and can be evaluated by the ratio between the hole carrier density ($n_h$) and the electron carrier density ($n_e$) [38]. According to the previous study [38], when the Fermi level is close to the Weyl points, $n_e/n_h$ is approximately 1.08. Hence, we measured $\bar{\gamma}_{yx}$ and $\bar{\gamma}_{xx}$ in more than ten samples and plot the absolute value of $\bar{\gamma}_{yx}$ and $\bar{\gamma}_{xx}$ at 1 T as a function of $n_e/n_h$ (see also [37] for the evaluation of $n_e$ and $n_h$). As shown in Figs. 4(a) and (b), $|\bar{\gamma}_{yx}|$ and $|\bar{\gamma}_{xx}|$ systematically depend on $n_e/n_h$ or equivalently the Fermi level position. Notably, both $|\bar{\gamma}_{yx}|$ and $|\bar{\gamma}_{xx}|$ change by more than four orders of magnitude and are largest when the Fermi level is close to the Weyl points. The observed substantial increase in the magnetochiral anisotropy responses around the Weyl points cannot be explained by the previously reported semiclassical calculation of the nonreciprocal responses in WTe$_2$ [20], in which an effect of the Berry curvature is not included; in this previous work, the ratio of the change in the calculated nonreciprocal response is less than 10 as a function of the Fermi level position.

The observed Fermi-level dependence of the magnetochiral anisotropy effect can be understood by a semiclassical treatment of current responses including the contribution from the Berry curvature. We use Boltzmann equation to derive momentum distribution under the electric field and the magnetic field and calculate the second-harmonic Hall and nonreciprocal magnetoresistance responses (for detail, see [37]). We find that the combination of $C_2$ rotation symmetry of WTe$_2$ around the $z$ direction and the time reversal symmetry leads to a constraint that the transverse nonreciprocity $\rho_{yx}^{2f}$ and longitudinal nonreciprocity $\rho_{xx}^{2f}$ is proportional to the square of the relaxation time $\tau$. The semiclassical treatment shows that such $\tau^2$ contribution to the second-harmonic Hall current arises from energy shift due to orbital magnetic moment and the momentum shift related to chiral anomaly. These effects are substantially enhanced when the Fermi energy is close to the Weyl point reflecting the diverging Berry curvature around the Weyl point. Therefore, as shown in Figs. 4(c) and (d), large second-harmonic Hall and nonreciprocal magnetoresistance responses appear when the Fermi energy is tuned close to the Weyl points, which is qualitatively consistent with the observed enhancement of the second-harmonic Hall effect and the nonreciprocal magnetoresistance. In particular, when the Fermi energy is located between the Weyl and anti-Weyl points, the contribution from the two nodes add up in a constructive way, leading to giant magnetochiral anisotropy.

Our theory can also explain the experimentally observed difference in the magnitudes of the second-harmonic Hall effect and the nonreciprocal magnetoresistance. As already shown in Figs. 1(d) and (e), the magnitude of the second-harmonic Hall effect is larger than that of the nonreciprocal magnetoresistance in sample F. Furthermore, all other samples also have the same relationship regardless of the current direction with respect to the crystalline axis; as presented in Figs. 4(a) and (b), the magnitude of the second-harmonic Hall effect is larger than that of the nonreciprocal magnetoresistance for each sample. In the theory, as can be seen from Eq. S11 [37], the second order current contains the $\boldsymbol{E}\cdot\boldsymbol{B}$ term that originates from the chiral anomaly [40, 41]. Since the $\boldsymbol{E}\cdot\boldsymbol{B}$ term is largest for $j \parallel B$ (the second-harmonic Hall configuration) and zero for $j \perp B$ (the nonreciprocal magnetoresistance configuration), the second-harmonic Hall effect is always larger than the nonreciprocal magnetoresistance, if the Fermi level position is the same. Furthermore, the present theory is also consistent with the experimentally observed temperature dependence of $\rho_{yx}^{2f}/\rho_{xx}$ and $\rho_{xx}^{2f}/\rho_{xx}$; both $\rho_{yx}^{2f}/\rho_{xx}$ and $\rho_{xx}^{2f}/\rho_{xx}$ increase with decreasing temperature [see Figs. 3(b) and (c)]. Since the second-harmonic Hall effect and the nonreciprocal magnetoresistance in our theory are proportional to the square of the relaxation time, they should increase with decreasing temperature because of the enhancement of the relaxation time $\tau$ at low temperatures.

To summarize, we observed the giant magnetochiral anisotropy effect when the Fermi level is close to the Weyl points. A semiclassical calculation of magnetochiral anisotropy responses indicates that the diverging Berry curvature at the Weyl points plays a crucial role in the observed enhancement of the magnetochiral anisotropy. Our finding demonstrates that the Berry curvature has a large impact on the magnetochiral anisotropy effect and is important for exploring large diode effects.

**Acknowledgments**

This works was supported by JST FOREST Program (Grant No. JPMJFR203H), JST PRESTO (Grant No. JPMJPR19L9), JST CREST (Grant No. JPMJCR19T3), JSPS KAKENHI (Grant Nos. 22H04464, 22H05449, 21K18890, 21H01794, 19H05600, and 19K14667), and Murata Science Foundation

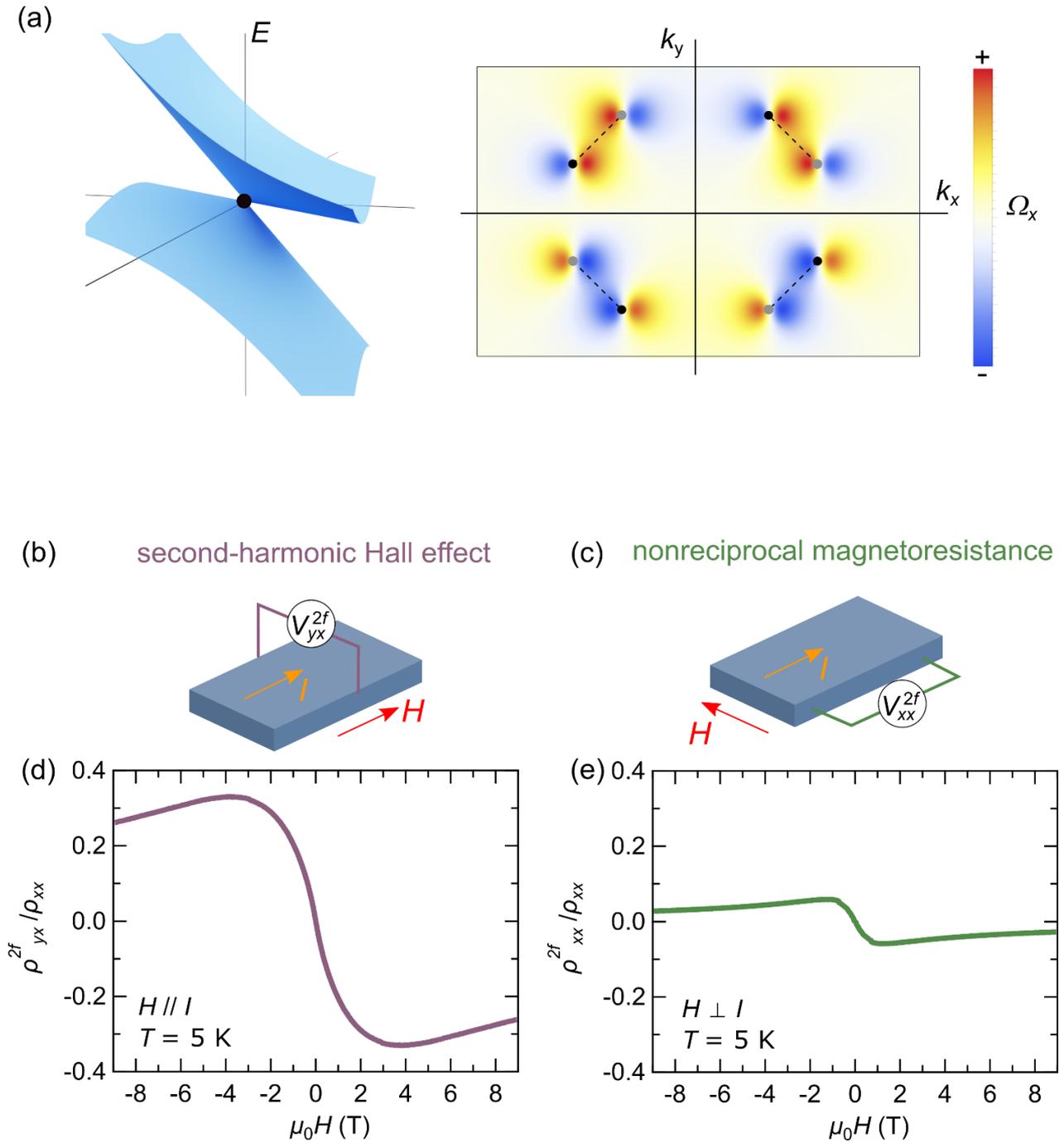

Fig. 1. (a) Schematics of the energy dispersion for a type-II Weyl semimetal near a Weyl point and the Berry curvature ($\Omega_x$) around the Weyl points in $WTe_2$. The black and grey circles represent Weyl points with the chirality $\eta = +1$ and -1, respectively. Experimental configurations for (b) the second-harmonic Hall effect and (c) the nonreciprocal magnetoresistance. Magnetic-field dependence of (d) the second-harmonic Hall effect and (e) the nonreciprocal magnetoresistance in sample F measured with $j = 8.8 \times 10^5$ $Am^{-2}$ and $f = 33$ Hz at 5 K. The current direction is parallel to the $b$-axis.

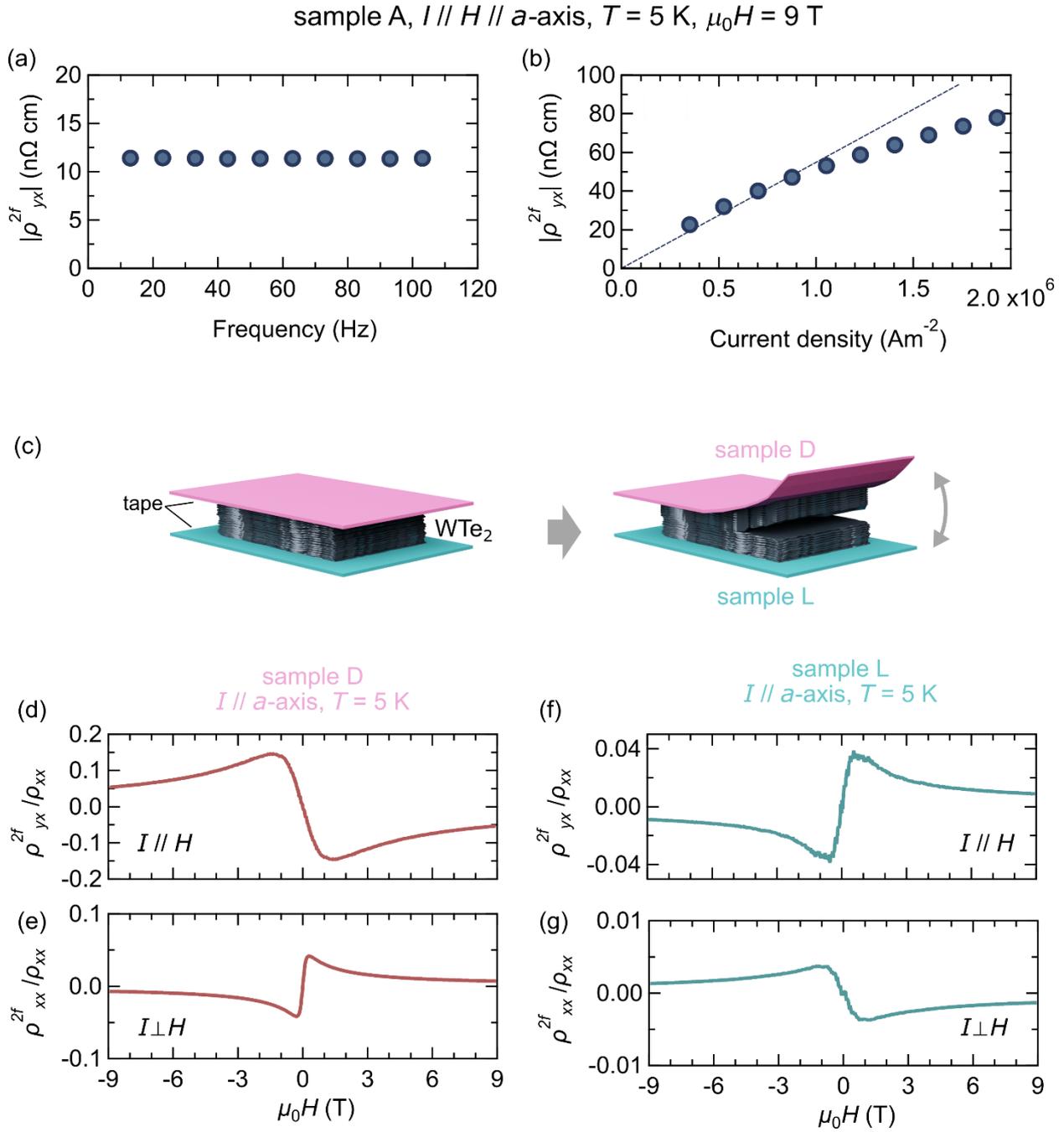

Fig. 2. (a) Frequency and (b) current density dependence of $\rho_{yx}^{2f}$ in sample A. The frequency dependence is measured with $j = 4.4 \times 10^6$ Am$^{-2}$ and the current dependence is measured with $f = 33$ Hz. The current direction is parallel to the $a$-axis. (c) Schematic of the fabrication of samples D and L, in which the polarity directions are opposite to each other. (d)-(g) Magnetic-field dependence of the second-harmonic Hall effect in sample D and sample L and the nonreciprocal magnetoresistance in sample D and sample L measured with $j = 1.2 \times 10^6$ Am$^{-2}$ and $f = 33$ Hz.

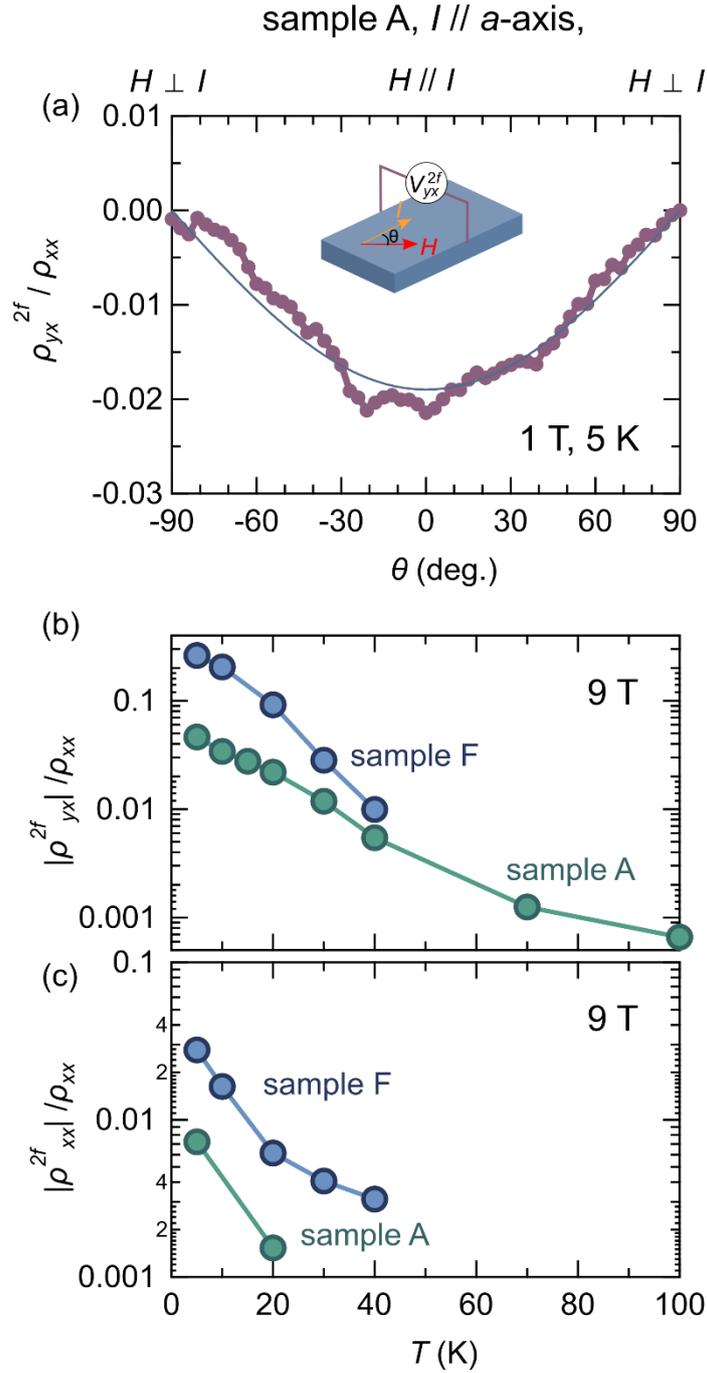

Fig. 3. (a) Angular dependence of $\rho_{yx}^{2f}/\rho_{xx}$ measured with $j = 8.8\times10^5$ Am$^{-2}$, $f = 3.3$ Hz, and $I \parallel a$-axis. The definition of $\theta$ is shown in the inset. The blue lines are fits to $\cos\theta$. (b), (c) Temperature dependence of $\rho_{yx}^{2f}/\rho_{xx}$ and $\rho_{xx}^{2f}/\rho_{xx}$ at 9 T in samples A and F. The current density for sample A and sample F is $j = 8.8\times10^5$ Am$^{-2}$ and $3.4\times10^5$ Am$^{-2}$, respectively. The current direction is parallel to the $a$-axis in sample A and the $b$-axis in sample F. The frequency of the current is $f = 33$ Hz in both samples.

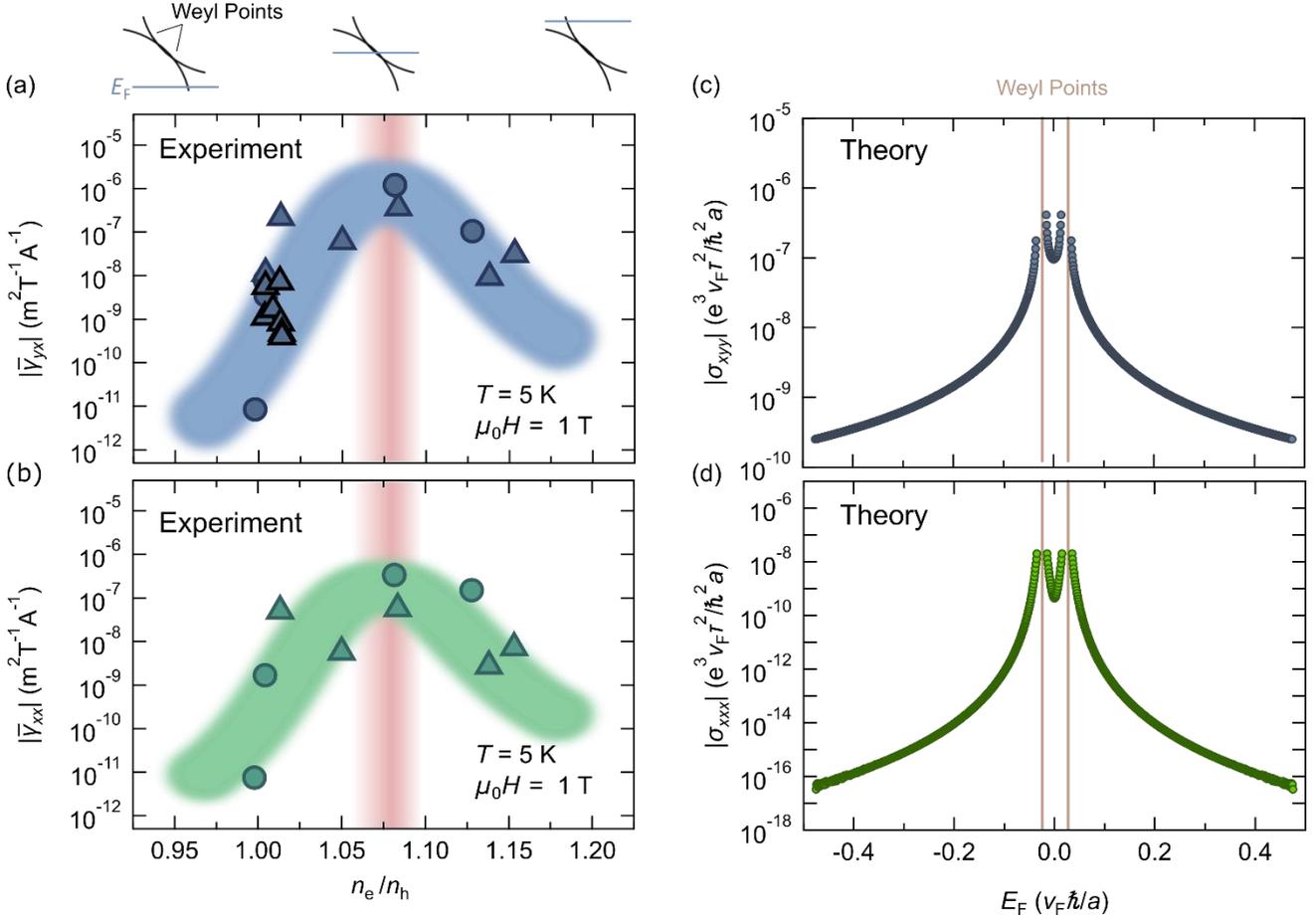

Fig. 4. (a), (b) Absolute values of $\bar{\gamma}_{yx} = 2\rho_{yx}^{2f}/\rho_{xx} jB$ and $\bar{\gamma}_{xx} = 2\rho_{xx}^{2f}/\rho_{xx} jB$ as a function of the ratio between electron carrier density ($n_e$) and hole carrier density ($n_h$). The circle and triangle points are samples measured with $I \parallel b$-axis and $I \parallel a$-axis, respectively. The upper panel schematically denotes the position of the Fermi level corresponding to the value of $n_e/n_h$. The region where the Fermi level is close to the Weyl points is denoted by the red shadow. The thick blue and green lines are a guide to the eyes. (c), (d) Absolute values of the second-harmonic Hall conductivity $\sigma_{xyy} = j_x/E_y^2$ and nonreciprocal magnetoresistance $\sigma_{xxx} = j_x/E_x^2$ as a function of the Fermi energy obtained from the semiclassical calculation. We set the magnetic field $B = 10^{-5}\ \hbar/ea^2$ and the energy separation of the two Weyl points $\Delta E = 0.05\ v_F\hbar/a$.

Supplemental Material for

# Giant Magnetochiral Anisotropy in Weyl-semimetal WTe$_2$ Induced by Diverging Berry Curvature

Tomoyuki Yokouchi[1], Yuya Ikeda[2], Takahiro Morimoto[2], and Yuki Shiomi[1]

[1]Department of Basic Science, The University of Tokyo, Tokyo 152-8902, Japan

[2]Department of Applied Physics, The University of Tokyo, Tokyo 113-8656, Japan

**Symmetry consideration**

The crystal structure of WTe$_2$ is orthorhombic with point group $C_{2v}$ and this material has also time reversal symmetry $T$ [S1]. Here we see what restrictions are imposed by such symmetries. The second-harmonic Hall effect can be represented as

$$j_x = \sigma_{xyy} E_y^2 B_y. \tag{S1}$$

By performing the symmetry operator $C_{2T} = C_{2z}T$, $E_y$ and $B_y$ are transformed to $-E_y$ and $B_y$, respectively. Therefore, the nonreciprocal Hall current $j_x$ must satisfy $j_x \rightarrow j_x$ under $C_{2T}$. In general, if we expand the current as $j = \Sigma_n j_n \tau^n$, where $\tau$ is the relaxation time, the expansion coefficients $j_n$ are converted to $(-1)^{n+1} j_n$ under $C_{2T}$. Since $j_0$ is the normal Hall current, we can see that the second-harmonic Hall current is proportional to $\tau^2$ at the lowest order.

The mirror symmetry in WTe$_2$ also constrains the direction of the nonreciprocal current response. For example, the nonreciprocal magnetoresistance (longitudinal nonreciprocal magneto-response) is not allowed when $B \parallel E$ where both $B$ and $E$ are normal to mirror plane (since the mirror operation leads to $J \rightarrow -J$, $E \rightarrow -E$ and $B \rightarrow B$). In contrast, the nonreciprocal magnetoresistance is allowed when $B \perp E$, which can be represented as

$$j_x = \sigma_{xxx} E_x^2 B_y. \tag{S2}$$

From the same argument as above, the nonreciprocal magnetoresistance is proportional to $\tau^2$ at the lowest order.

**Formalism of semiclassical calculation**

We use Boltzmann equation to derive momentum distribution under the electric field and the magnetic field and calculate the nonlinear current response. The semiclassical equation of motion for an electron under the electric field and the magnetic field are [S2, S3]

$$\dot{r} = \frac{1}{\hbar D}\left[\nabla_k \epsilon_k + eE \times \Omega_k + \frac{e}{\hbar}(\nabla_k \epsilon_k \cdot \Omega_k)B\right] \tag{S3}$$

$$\hbar \dot{k} = \left[-eE - \frac{e}{\hbar}\nabla_k \epsilon_k \times B - \frac{e^2}{\hbar}(E \cdot B)\Omega_k\right] \tag{S4}$$

$$D = 1 + \frac{e}{\hbar}B \cdot \Omega_k. \tag{S5}$$

The third term in $\dot{k}$ describe the momentum shift due to the chiral anomaly within semiclassical approach. The energy dispersion $\epsilon_k$ includes correction due to magnetic orbital moment,

$$\epsilon_k = \epsilon_k^0 - m_k \cdot B. \tag{S6}$$

In the uniform system, current density $j$ is given by

$$j = -e \int [dk] D\dot{r} f, \tag{S7}$$

where $f$ is the distribution function and we define $[dk] \equiv dk/(2\pi)^3$.

Now, we expand the distribution function with respect to $E$ as

$$f = f_0 + f_1 + f_2, \tag{S8}$$

where $f_0 = \theta(E_F - \epsilon_k^0 + m_k \cdot B)$ is the unperturbed distribution function. The steady state is determined by the Boltzmann equation with relaxation time approximation

$$\frac{df}{dt} = \dot{k} \cdot \nabla_k f = \frac{f_0 - f}{\tau}. \tag{S9}$$

The second-order response proportional to $\tau^2$ is obtained from $f_2$. Substituting Eq. (S8) successively into Eq. (S9), $f_2$ is given by

$$f_2 = \left\{\frac{e\tau}{\hbar D}\left[E + \frac{e}{\hbar}(E \cdot B)\Omega_k\right] \cdot \nabla_k\right\}^2 f_0, \tag{S10}$$

where we have dropped the term involving $\nabla_k \epsilon_k \times B$ because this term is perpendicular to $\nabla_k f_i = (\nabla_k \epsilon_k)\partial_\epsilon f_i$ $(i = 0,1)$. The second-order current response (of the frequency $2\omega$) is obtained as

$$j_2 = -e \int [dk] D\dot{r} f_2$$

$$= -\tau^2 \frac{e^3}{\hbar^2} \int [dk] \left[v_k + \frac{e}{\hbar}(v_k \cdot \Omega_k)B\right] \left\{\frac{e\tau}{\hbar D}\left[E + \frac{e}{\hbar}(E \cdot B)\Omega_k\right] \cdot \nabla_k\right\}^2 f_0, \tag{S11}$$

with $\boldsymbol{v}_{\boldsymbol{k}} = (1/\hbar)\boldsymbol{\nabla}_{\boldsymbol{k}}\epsilon_{\boldsymbol{k}}$. Now we focus on the second-harmonic Hall current $j_x = \sigma_{xyy}E_y^2 B_y$ and thus drop the term involving $(\boldsymbol{v}_{\boldsymbol{k}} \cdot \boldsymbol{\Omega}_{\boldsymbol{k}})\boldsymbol{B}$. Integrating by parts leads to

$$j_x = \tau^2 \frac{e^3}{\hbar^2} E_y^2 \int_{\mathrm{BZ}} [d\boldsymbol{k}] \frac{1}{D}\left(\frac{\partial(v_{\boldsymbol{k}}^x/D)}{\partial k_y} + \frac{e}{\hbar}B_y \boldsymbol{\nabla}_{\boldsymbol{k}} \cdot (v_{\boldsymbol{k}}^x \boldsymbol{\Omega}_{\boldsymbol{k}}/D)\right)\left[\boldsymbol{e}_y + \frac{e}{\hbar}B_y \boldsymbol{\Omega}_{\boldsymbol{k}}\right] \cdot \boldsymbol{\nabla}_{\boldsymbol{k}} f_0$$

$$= \tau^2 \frac{e^3}{\hbar^2} E_y^2 \int_{\mathrm{BZ}} [d\boldsymbol{k}] \frac{1}{D}\left(\frac{\partial(v_{\boldsymbol{k}}^x/D)}{\partial k_y} + \frac{e}{\hbar}B_y \boldsymbol{\nabla}_{\boldsymbol{k}} \cdot (v_{\boldsymbol{k}}^x \boldsymbol{\Omega}_{\boldsymbol{k}}/D)\right)\left[v_{\boldsymbol{k}}^y + \frac{e}{\hbar}B_y \boldsymbol{v}_{\boldsymbol{k}} \cdot \boldsymbol{\Omega}_{\boldsymbol{k}}\right]\frac{\partial f_0}{\partial \epsilon}$$

$$= \tau^2 \frac{e^3}{\hbar^2} E_y^2 \int_S [d\boldsymbol{k}] \frac{1}{D|\boldsymbol{v}_{\boldsymbol{k}}|}\left(\frac{\partial(v_{\boldsymbol{k}}^x/D)}{\partial k_y} + \frac{e}{\hbar}B_y \boldsymbol{\nabla}_{\boldsymbol{k}} \cdot (v_{\boldsymbol{k}}^x \boldsymbol{\Omega}_{\boldsymbol{k}}/D)\right)\left[v_{\boldsymbol{k}}^y + \frac{e}{\hbar}B_y \boldsymbol{v}_{\boldsymbol{k}} \cdot \boldsymbol{\Omega}_{\boldsymbol{k}}\right]. \quad (S12)$$

In the third line, the range of integration is changed to the Fermi surface $S$ because $\partial_\epsilon f_0 = \delta(E_\mathrm{F} - \epsilon)$. The second-harmonic Hall conductivity shows a rapid enhancement around the Weyl points reflecting the diverging Berry curvature. We note that the second-harmonic Hall response from the Berry curvature dipole mechanism arises from $f_1$, which is in the linear order in $\tau$ and is prohibited in the present setup by $C_{2T}$ symmetry for WTe$_2$.

From Eq. (S11), we can also obtain the formula for the nonreciprocal magnetoresistance $j_x = \sigma_{xxx}E_x^2 B_y$. Since $\boldsymbol{B}$ and $\boldsymbol{E}$ are perpendicular, we can drop the chiral anomaly terms involving $(\boldsymbol{E} \cdot \boldsymbol{B})$ in the distribution function under the electromagnetic field, which leads to

$$j_x = \tau^2 \frac{e^3}{\hbar^2} E_x^2 \int_{\mathrm{BZ}} [d\boldsymbol{k}] \frac{1}{D}\frac{\partial(v_{\boldsymbol{k}}^x/D)}{\partial k_x}\frac{\partial f_0}{\partial k_x}$$

$$= \tau^2 \frac{e^3}{\hbar^2} E_y^2 \int_S [d\boldsymbol{k}] \frac{v_{\boldsymbol{k}}^x}{D|\boldsymbol{v}_{\boldsymbol{k}}|}\frac{\partial(v_{\boldsymbol{k}}^x/D)}{\partial k_x}. \quad (S13)$$

This nonreciprocal current response is nonzero due to the orbital magnetic moment. However, as discussed in the main text, the second-harmonic Hall response includes not only the orbital magnetic moment effect but also the chiral anomaly effect $(\boldsymbol{E} \cdot \boldsymbol{B})$, so the magnitude of the second-harmonic Hall effect is larger than the nonreciprocal magnetoresistance.

### Second-harmonic Hall conductivity of tilted Weyl semimetals

We consider the Weyl Hamiltonian with a tilt,

$$H = \eta v_\mathrm{F} \boldsymbol{k} \cdot \boldsymbol{\sigma} + v_\mathrm{t} k_\mathrm{t} \sigma_0, \quad (S14)$$

where $k_t = k_x \cos\theta_t + k_y \sin\theta_t$ is the direction of tilt and $\eta = \pm 1$ specifies the chirality. The energy dispersion is given by

$$\epsilon_k^\pm = \pm v_F k + v_t k_t - \boldsymbol{m_k} \cdot \boldsymbol{B}, \tag{S15}$$

which is linear in $k$. For the above the Hamiltonian, the Berry curvature and the orbital magnetic moment are written as

$$\boldsymbol{\Omega_k^\pm} = \mp \eta \frac{\boldsymbol{k}}{2k^3}, \tag{S16}$$

$$\boldsymbol{m_k} = -\eta \frac{\boldsymbol{k}}{2k^2}. \tag{S17}$$

WTe$_2$ host 4 pairs of Weyl and anti-Weyl fermions which are connected with mirror symmetries along the $x$ and $y$ directions as shown in Fig. S1(a). Explicitly, four Weyl points with the tilt angle and the chirality $(\theta_t, \eta), (-\theta_t, -\eta), (\pi + \theta_t, \eta), (\pi - \theta_t, -\eta)$ are connected with mirror symmetries. In the magnetic field, the Fermi surface of each Weyl point shifts due to the orbital magnetization as illustrated in Fig. S1(b). Using Eq. (S12), the second-harmonic Hall conductivity is computed for these four pairs of Weyl fermions as shown in Fig. S1(c). Due to the divergence of the Berry curvature at each of the Weyl points at different energies, the second-harmonic Hall conductivity shows a rapid enhancement around the Weyl points. In particular, when the Fermi energy is located between the two Weyl nodes, the contributions from the two Weyl nodes add up, leading to a large second-harmonic Hall effect shown as Figure 4(c) in the main text. We note that, in this calculation, the integration is performed only on the Fermi surfaces that change continuously from the ones in the absence of the magnetic field. The magnetic field dependence of the second-harmonic Hall conductivity is also calculated as shown in Fig. S1(d). Only $B$-odd terms are allowed due to $C_2$ symmetry along the $z$ direction. We see that the conductivity is $B$-linear in the small magnetic field region, which corresponds to the nonreciprocal second-harmonic Hall current in Eq. (S1). These results show that the experimentally observed Fermi-level dependence can be explained by a semiclassical calculation of nonreciprocal responses which incorporates the effect of energy shift due to the orbital magnetic moment and momentum shift induced by the chiral anomaly of Weyl fermions.

We note that the unit of the energy in Fig. 1 is given by $[v_F \hbar/a]$. For typical parameters for Weyl semimetals, $v_F = 10^5$ m/s and $a = 3$Å, this energy unit amounts to $v_F \hbar/a \simeq 0.2$ eV, where the Weyl node separation of $\Delta E = 0.05\ v_F \hbar/a \simeq 0.01$ eV that is used for Fig. S1(c) becomes comparable with that for WTe$_2$ [S1].

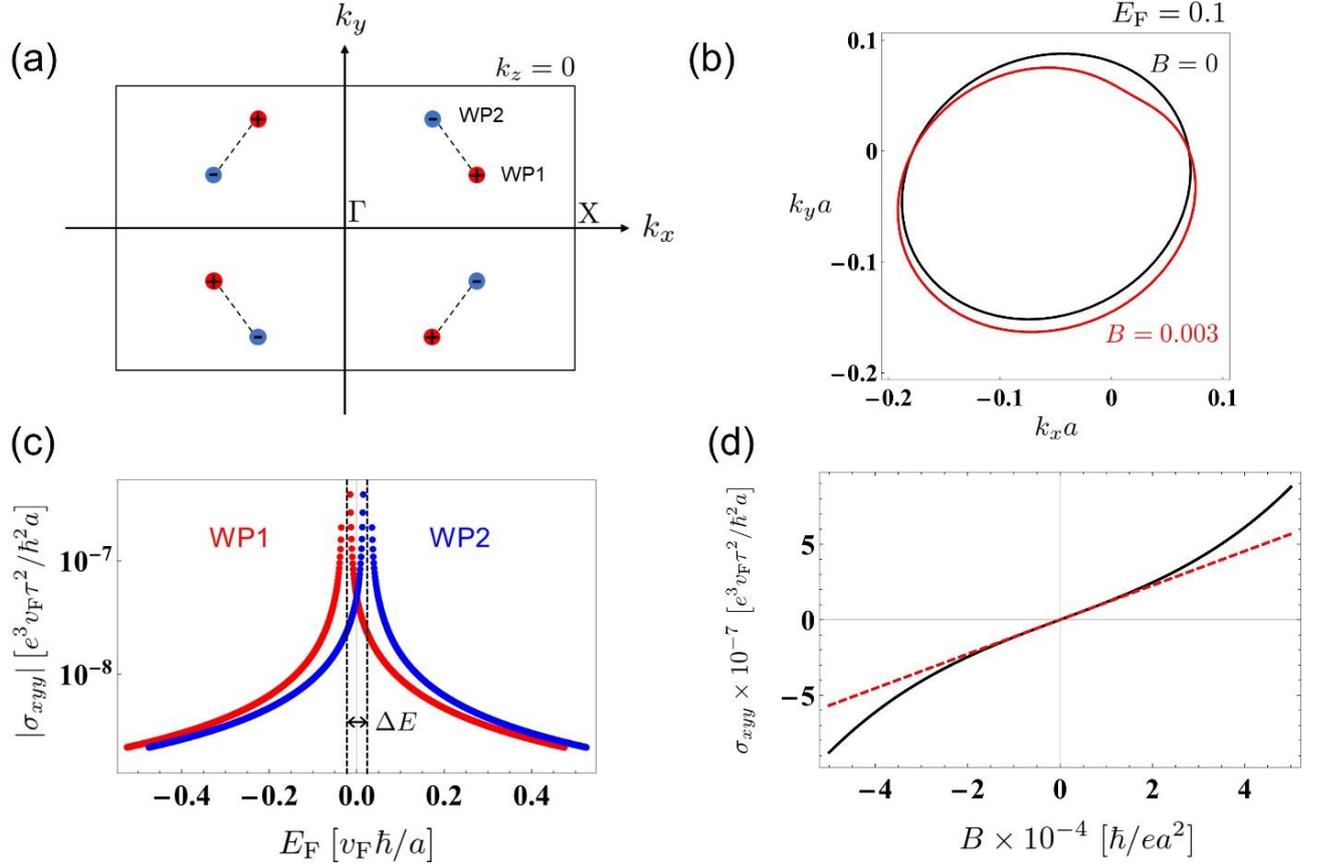

Fig. S1. Theoretical calculation of second-harmonic Hall conductivity of tilted Weyl fermions. (a) Schematic picture of four pairs of Weyl and anti-Weyl points in WTe$_2$, where their chirality is marked with red ($\eta = 1$) and blue ($\eta = -1$). The four Weyl pairs appear at the $k_z = 0$ plane in a way satisfying the mirror symmetries along $x$ and $y$ directions. (b) Cross-section of Fermi surface at the $k_z = 0$ plane in the absence (black line) and the presence (red line) of the magnetic field $B$. In the presence of magnetic field in the $y$ direction, the Fermi surface shifts down in the $k_y$ direction due to the orbital magnetization. ($E_F$ and $B$ are expressed in a dimensionless unit with $e = \hbar = a = 1$ for simplicity.) (c) The Fermi level dependence of the second-harmonic Hall conductivity of two Weyl points with energy $\pm \Delta E/2$. We set the magnetic field $B = 10^{-5}\ \hbar/ea^2$ and the Weyl node separation $\Delta E = 0.05\ v_F \hbar/a$. The second-harmonic Hall conductivity shows a rapid enhancement around the Weyl points reflecting the diverging Berry curvature. (d) The magnetic field dependence of the second-harmonic Hall conductivity (black line). The conductivity is $B$-odd and, in the small field region, is $B$-linear (the red line is linear-fitting). We used the parameters, $v_t/v_F = 0.5$, $\theta_t = \pi/3$.

## Sample preparation

Single crystals of $WTe_2$ were synthesized by a self-flux method [S4]. First, the mixture of high-purity W and Te powders was heated up to 835 °C in an evacuated quartz tube. Next, the tube was cooled down to 535 °C at the rate of 2 °C/hour. Then, Te was removed by centrifuging. Finally, we annealed the crystals at 415 °C in an evacuated quartz tube for several days.

The typical size of obtained samples is $2\times0.3\times0.04$ mm$^3$. In Fig. S2 and S3, we show the temperature dependence of the resistivity and the magnetoresistance, both of which are consistent with the previous report [S5].

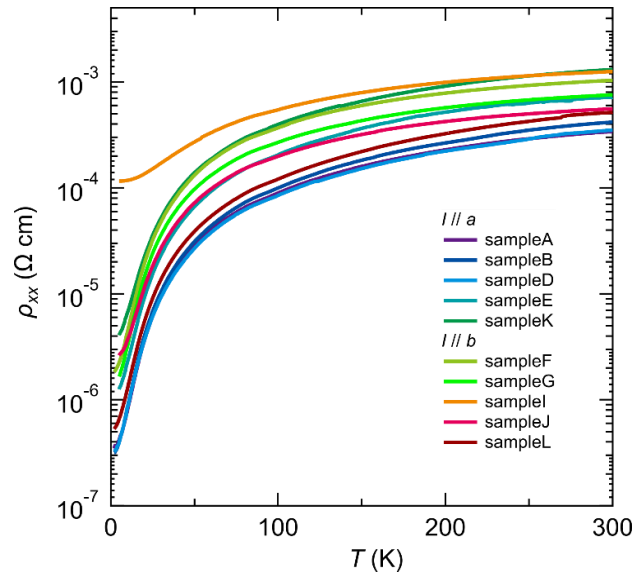

Fig. S2. Temperature dependence of longitudinal resistivity for sample A to sample L.

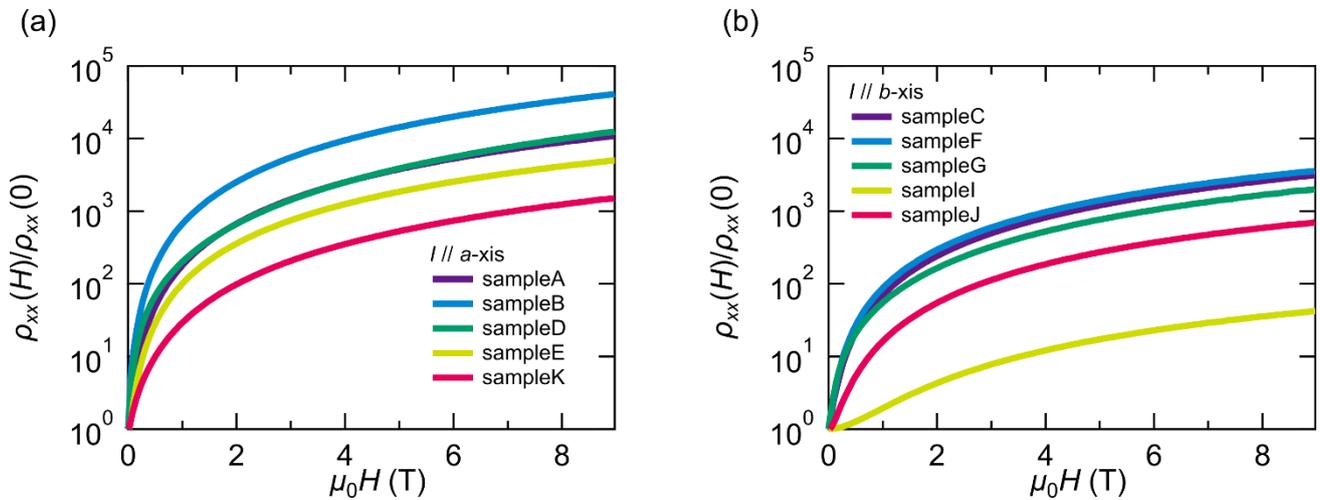

Fig. S3. Magnetic field dependence of longitudinal resistivity for samples with $I \parallel a$-axis (a) and with $I \parallel b$-axis (b) at 5 K. The magnetic field is applied parallel to the $c$-axis.

**Transport measurements**

We connected gold wires to WTe$_2$ single crystals with silver paste (Dupont 4922 or Epo-Tek H20E). The electrode distances and the thickness of the samples were measured with a laser microscope (Keyence VK-9500). Temperature and magnetic-filed dependence of longitudinal resistivity and Hall resistivity (Figs. S2-S4) were measured by using dc-transport option of Physical Property Measurement System (PPMS).

Second harmonic resistivity was measured by using a lock-in technique (NF LI5650) and AC current source (Lake Shore 155); we input low-frequency ($f$) ac current and measured second harmonic voltage. The second harmonic longitudinal and transverse resistivities ($\rho_{xx}^{2f}$ and $\rho_{yx}^{2f}$) are defined as $\rho_{xx}^{2f} = V_{xx}^{2f} S/Id$ and $\rho_{yx}^{2f} = V_{yx}^{2f} S/Id$, respectively. Here, $S$, $I$, $d$, $V_{xx}^{2f}$, and $V_{yx}^{2f}$ are cross-section area of WTe$_2$, the current, the distance between voltage terminal, the longitudinal second harmonic voltage, and the transverse second harmonic voltage, respectively. Here, $\rho_{xx}^{2f}$ and $\rho_{yx}^{2f}$ were anti-symmetrized against the magnetic field as follows: $\rho_{xx}^{2f}(H) = [\rho_{xx,\text{meas.}}^{2f}(H) - \rho_{xx,\text{meas.}}^{2f}(-H)]/2$ and $\rho_{yx}^{2f}(H) = [\rho_{yx,\text{meas.}}^{2f}(H) - \rho_{yx,\text{meas.}}^{2f}(-H)]/2$, where $\rho_{xx,\text{meas.}}^{2f}(H)$ and $\rho_{yx,\text{meas.}}^{2f}(H)$ are the measured second harmonic longitudinal and transverse resistivities, respectively. We show $\rho_{xx,\text{meas.}}^{2f}(H)$ and $\rho_{yx,\text{meas.}}^{2f}(H)$ for sample F in Fig. S5; the antisymmetric component is dominant in $\rho_{xx,\text{meas.}}^{2f}(H)$ and $\rho_{yx,\text{meas.}}^{2f}(H)$. In the calculation of $\rho_{xx}^{2f}/\rho_{xx}$ and $\rho_{yx}^{2f}/\rho_{xx}$, $\rho_{xx}$ simultaneously measured together with $\rho_{xx}^{2f}$ or $\rho_{yx}^{2f}$ with a lock-in amplifier is used. The figures of merit of the second-harmonic Hall effect $\bar{\gamma}_{yx}$ and nonreciprocal magnetoresistance $\bar{\gamma}_{xx}$ are defined as $\bar{\gamma}_{yx} = 2\rho_{yx}^{2f}/(\rho_{xx}jB)$ and $\bar{\gamma}_{xx} = 2\rho_{xx}^{2f}/(\rho_{xx}jB)$, where $j$ is the current density [S6]. We show $\rho_{xx}$, $\rho_{yx}^{2f}$, $\rho_{yx}^{2f}/\rho_{xx}$, $\rho_{xx}^{2f}$, $\rho_{xx}^{2f}/\rho_{xx}$ for all samples in Fig. S6 and Fig. S7. In the measurement of the angular dependence [Fig. 3(a) in the main text], we anti-symmetrized $\rho_{yx}^{2f}$ with respect to the magnetic field as well as the angle as follows: $\rho_{yx}^{2f}(\theta) = \{[(\rho_{yx,\text{meas.}}^{2f}(\theta,H) - \rho_{yx,\text{meas.}}^{2f}(\theta,-H))] - [\rho_{yx,\text{meas.}}^{2f}(\theta+\pi,H) - \rho_{yx,\text{meas.}}^{2f}(\theta+\pi,-H)]\}/4$.

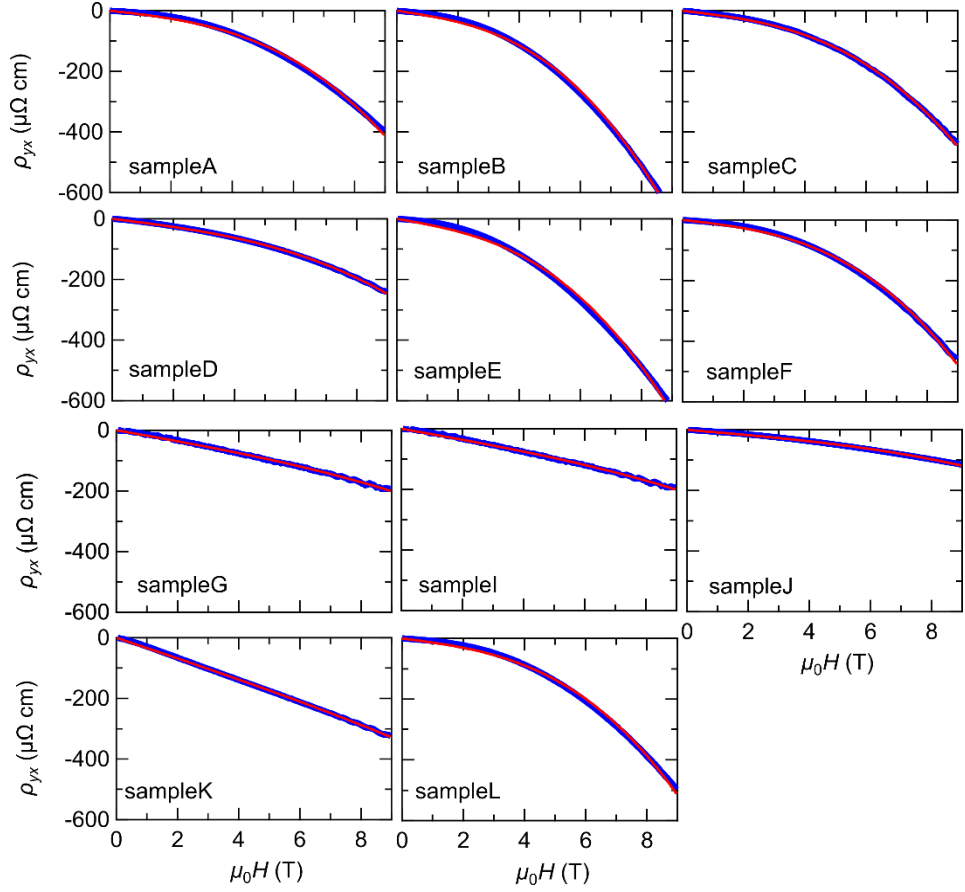

Fig. S4. Magnetic field dependence of Hall resistivity at 5 K. The blue curves show the measured Hall resistivity, and the red curves are fitting curves with the two-carrier model [Eq. (1) in the main text]. The magnetic field is applied parallel to the *c*-axis.

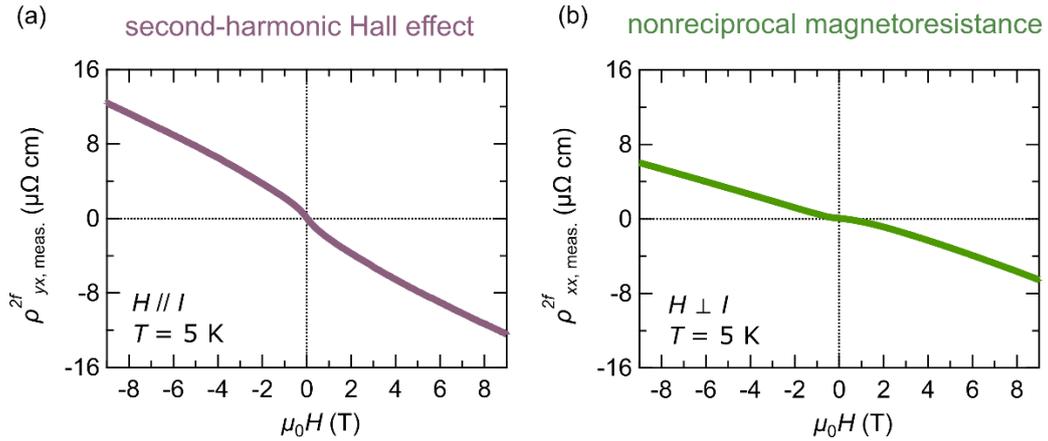

Fig. S5. Magnetic-field dependence of the raw data of second-harmonic Hall effect (a) and the nonreciprocal magnetoresistance (b) in sample F measured with $j = 8.8 \times 10^5$ Am$^{-2}$ and $f = 33$ Hz at 5 K. The current direction is parallel to the *b*-axis. The corresponding anti-symmetrized data normalized by $\rho_{xx}$ is shown in Fig. 1 in the main text.

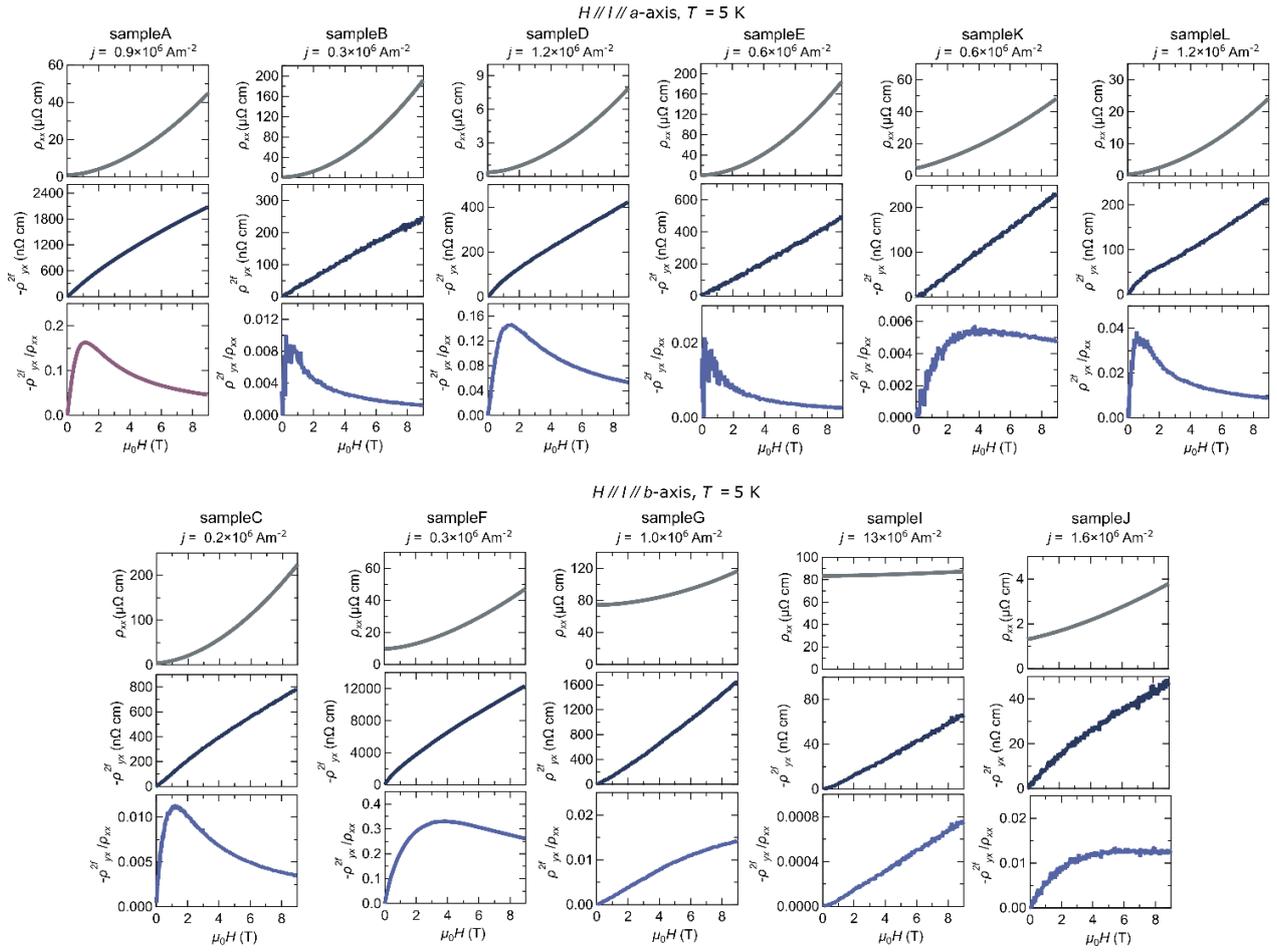

Fig. S6. Magnetic field dependence of longitudinal resistivity, second-harmonic Hall resistivity, and second-harmonic Hall resistivity normalized by the longitudinal resistivity measured with $H \parallel I$ at 5 K.

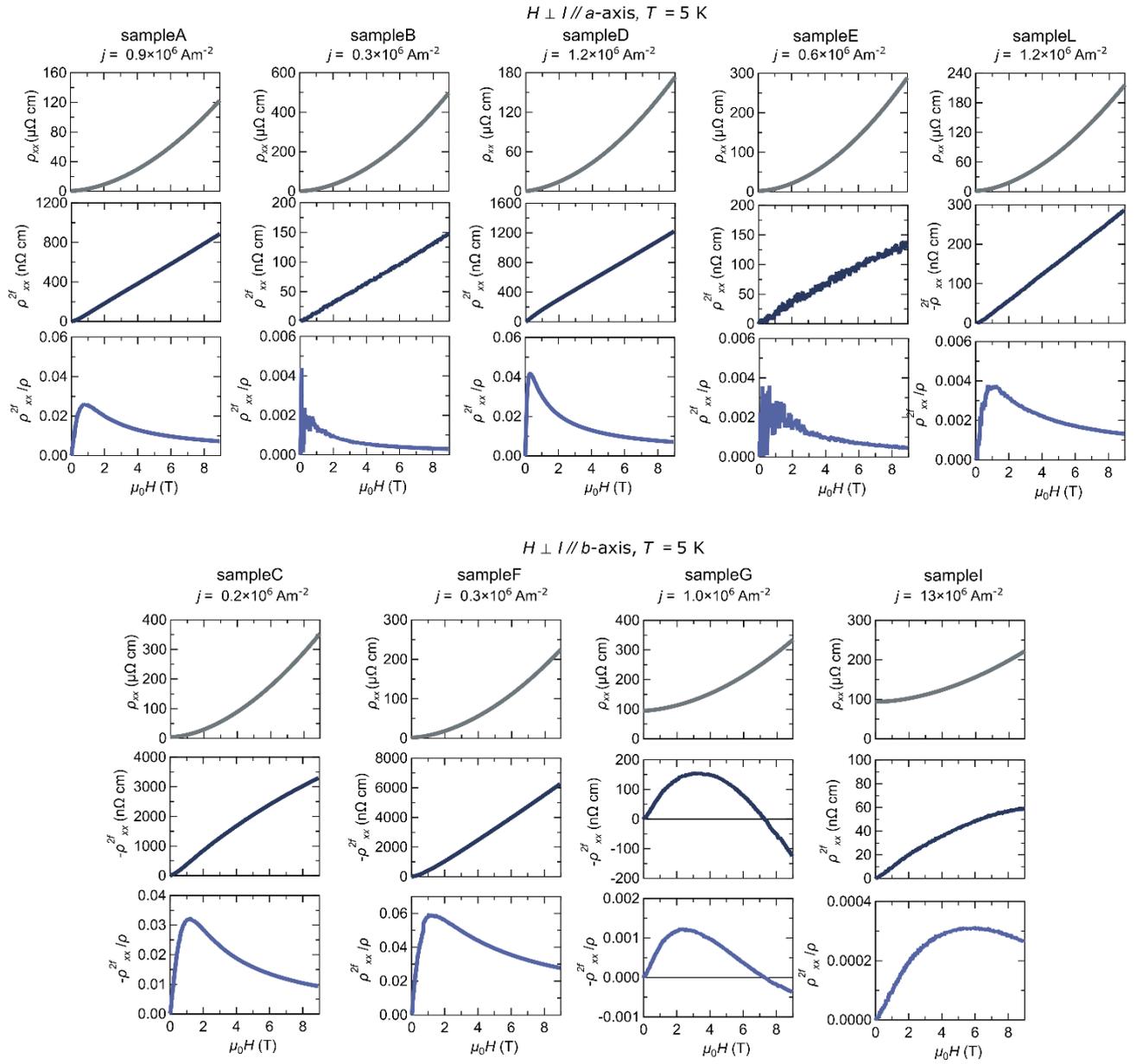

Fig. S7. Magnetic field dependence of longitudinal resistivity, nonreciprocal magnetoresistance, and nonreciprocal magnetoresistance normalized by the longitudinal resistivity measured with $H \perp I$ at 5 K.

**Evaluation of hole and electron carrier densities**

The hole and electron carrier densities ($n_h$ and $n_e$) are determined by adopting a two-band model with one-electron band and one-hole band as follows [S8]:

$$\rho_{yx} = \frac{(n_h\mu_h^2 - n_e\mu_e^2) + (n_h - n_e)\mu_h^2\mu_e^2 B^2}{(n_h\mu_h + n_e\mu_e)^2 + (n_h - n_e)^2\mu_h^2\mu_e^2 B^2}\frac{B}{e}, \quad (S18)$$

where $n_h, n_e, \mu_h, \mu_e, e$ are hole carrier density, electron carrier density, hole mobility, electron mobility, and elementary charge, respectively. In Fig. S4, we show the magnetic-field dependence of Hall resistivity and fitting curves with Eq. (S18). The Hall resistivity is well fitted with the two-band model.

In bulk WTe$_2$, the carrier density is controlled by annealing because of a change in the Te composition [S9]. In sample K, we repeated the transport measurement and the annealing to vary the Te composition. Each annealing was performed at 150 °C for approximately 1 hour in a vacuum. We performed the annealing and the measurement of the second-harmonic Hall effect seven times in sample K. After the repeated experiments, $n_h/n_e$ changes by 0.97 %, eventually.